\documentclass[11pt]{llncs}
\usepackage{latexsym}
\usepackage[pdf]{pstricks}
\usepackage{pstricks-add}
\usepackage{auto-pst-pdf}
\usepackage{pst-node}
\usepackage{pst-tree}
\usepackage{amssymb}
\usepackage{amsmath,bm}
\usepackage{stmaryrd}
\usepackage{xspace}
\usepackage{fullpage}

\newcommand{\e}[1]{{\bf #1}}
\newcommand{\logic}[1]{\textsc{#1}\xspace}
\newcommand{\FOL}{\logic{Fo}}
\newcommand{\domain}{\Delta^{\inte}}
\newcommand{\Var}[1]{\textit{Var}(#1)}
\newcommand{\g}[1]{\logic{#1}}
\newcommand{\inte}{\f{I}}
\newcommand{\tup}[1]{(#1)}
\newcommand{\card}[1]{\#(#1)}
\newcommand{\set}[1]{\{#1\}}
\newcommand{\f}[1]{\mathcal{#1}}

\newcommand{\LogSpace}{\textsc{LogSpace}\xspace}
\newcommand{\PTime}{\textsc{PTime}\xspace}
\newcommand{\NP}{\textsc{NP}\xspace}
\newcommand{\ex}{\hspace*{\fill}$\clubsuit$}
\newcommand{\dlliterc}{\g{DL-Lite}_{R,\sqcap}\xspace}
\renewcommand{\vec}[1]{\bar{#1}} % renewing vector notation
\title{The Expressive Power of $\dlliterc$}
\author{
Camilo Thorne
}
\institute{
\begin{tabular}{c}
IBM CAS Trento - Trento RISE\\
Piazza Manci 17\\
38123, Povo di Trento (Italy)\\
{\tt c.thorne.email@trentorise.eu}
\end{tabular}
}
\begin{document}
%%%%%%%%%%%%%%%%%%%%%%%%%%%%%%%%%%%%%%%%%%%%%%%%%%%%%%%%%%%%%%%

\maketitle

%%%%%%%%%%%%%%%%%%%%%%%%%%%%%%%%%%%%%%%%%%%%%%%%%%%%%%%%%%%%%%%

\begin{abstract}
Description logics are knowledge representation formalisms
that provide the formal underpinning of the semantic 
web and in particular of the OWL web ontology language.
In this paper we investigate the
expressive power of $\dlliterc$, and some of
its computational properties. We rely on simulations
to characterize the absolute expressive power of $\dlliterc$
as a {\em concept language}, and to show that disjunction is not expressible.
We also show that no simulation-based closure property exists
for $\dlliterc$ assertions. Finally, we show
that query answering of unions of conjunctive queries is
\NP-complete.
\end{abstract}

%%%%%%%%%%%%%%%%%%%%%%%%%%%%%%%%%%%%%%%%%%%%%%%%%%%%%%%%%%%%%%%
\section{Introduction}\label{intro}
%%%%%%%%%%%%%%%%%%%%%%%%%%%%%%%%%%%%%%%%%%%%%%%%%%%%%%%%%%%%%%%

Description logics (DLs) are knowledge representation formalisms
that provide the formal underpinning of the semantic 
web and in particular of the OWL web ontology language\footnote{http://www.w3.org/TR/owl-features/}.
In this paper we are interested in investigating the
expressive power of the DL known as \g{DL-Lite}$_{R,\sqcap}$
\cite{Calvanese2006B}. The \g{DL-Lite} family of logics, of which
$\dlliterc$ makes part,
has been proposed by Calvanese et al. as a foundation of
ontology-based data access systems. They are intended
\cite{Calvanese2005A,Calvanese2007D}
as the least expressive DLs capable of capturing the main
features of conceptual modelling languages such as UML\footnote{http://www.omg.org/uml/}.
By the expressive power of a DL we understand
\textit{(i)} the
computational complexity of its reasoning problems
and \textit{(ii)} its model-theoretic properties.
As most DLs, $\dlliterc$ 
is contained in $\FOL^2$, the 2-variable
fragment of \FOL and is therefore decidable
\cite{Borgida1996,Hudstadt2004,DLHandbook}. However,
its expressive power is still not known completely.

DLs model
domains in terms
of concepts (representing classes of objects), 
and binary relations known as roles (representing relations and
attributes of objects) \cite{DLHandbook}, all of which
are structured
into hierarchies by concept and role inclusion assertions.
Extensional information (the data), by contrast,
is conveyed by membership assertions. 
This information can be accessed by posing suitable \FOL
formulas, viz., unions of conjunctive queries.
This crucial reasoning problem is known as the knowledge
base query answering problem.

The main contributions of this paper consist, on the one hand,
in determining the (so-called) combined complexity of $\dlliterc$'s
query answering problem and, on the other hand,
to define what
we call $\dlliterc$ simulations.
This relation stems from the notion of bisimulations
(see e.g. \cite{vanBenthem2006}) for modal logics, known
to hold for the DL $\f{ALC}$ \cite{DLHandbook}, that
has been proposed \cite{deRijke1997} as a means
of characterizing the (absolute)
expressivity of arbitrary DLs as concept languages.

The structure of this paper is as follows.
Section~\ref{two} recalls \textit{(i)} $\dlliterc$'s syntax
and semantics and \textit{(ii)} those of unions of conjunctive queries.
In section~\ref{two} we characterize the 
combined complexity of answering unions of conjunctive queries
over $\dlliterc$ knowledge bases.
In section~\ref{three} we introduce the notion
of $\dlliterc$ simulations
and show that a \FOL formula is equivalent to a $\dlliterc$
concept when and only when it is closed under $\dlliterc$ simulations.
In section~\ref{four} we show that no such closure property
exists for assertions.
Finally, in section~\ref{five} we sum up our conclusions.

%%%%%%%%%%%%%%%%%%%%%%%%%%%%%%%%%%%%%%%%%%%%%%%%%%%%%%%%%%%%%%%
\section{Preliminaries}\label{one}
%%%%%%%%%%%%%%%%%%%%%%%%%%%%%%%%%%%%%%%%%%%%%%%%%%%%%%%%%%%%%%%

The syntax of
$\dlliterc$ is defined by the grammar:
\begin{itemize}
\item $R ::= P \mid P^-$,
\item $D ::= A \mid \exists R \mid D \sqcap D'$ (left concepts),
\item $E ::= C \mid \neg A \mid \neg \exists R \mid \exists R . E$ (right concepts),
\end{itemize}
where $A$ stands for an atomic concept symbol (a unary predicate),
$P$ for an atomic role symbol (a binary predicate) and $R^-$ for its inverse.

Concepts combine into {\em concept inclusion assertions} of the form
$D \sqsubseteq E$, where $D$ is a left concept, $E$ is a right concept
and $\sqsubseteq$ is the {\em subsumption} relation. Roles into
{\em role inclusion assertions} of the form $R \sqsubseteq R'$.
A {\em teminology} $\f{T}$ (TBox) is a set of such assertions.
A {\em membership assertion} is an assertion of the form 
$A(c)$ or $P(c,c')$, where $c,c'$ are object (or individual) constants.
We denote $\f{A}$ any set of membership assertions (ABox).
The integer
$\card{\f{A}}$ denotes the number of (distinct)
tuples occuring among the atoms in $\f{A}$.
The integer $\card{\f{T}}$ the number of axioms in the terminology.
A {\em knowledge base} is a pair $\tup{\f{T},\f{A}}$.

Let $\e{Dom}$ denote a countable infinite set of constants.
The semantics of $\dlliterc$
is based on \FOL\xspace {\em interpretations} $\inte := 
\tup{\domain, .^{\inte}}$, where
$\domain \subsetneq \e{Dom}$ is a non-empty {\em domain}.
Interpretations map each constant $c$ to itself,
each atomic concept $A$ to $A^{\inte} \subseteq \domain$
and each atomic role $P$ to $P^{\inte} \subseteq \domain \times \domain$
such that the following conditions hold:
\begin{itemize}
\item $(P^-)^{\inte} := \set{\tup{d, e} \in \domain \times \domain \mid 
\tup{e, d} \in P^{\inte}}$,
\item $(\exists R)^{\inte} := \set{d \in \domain \mid \text{exists } e \in \domain \text{ s.t. } 
\tup{d, e} \in R^{\inte}}$, 
\item $(D \sqcap D')^{\inte} := D^{\inte} \cap D'^{\inte}$,
\item $(\neg A)^{\inte} := \domain - A^{\inte}$,
\item $(\neg \exists R)^{\inte} := \domain - (\exists R)^{\inte}$, and
\item $(\exists R . E)^{\inte} := \set{d \in \domain 
\mid \text{exists } e \in \domain \text{ s.t. } \tup{d, e}
\in R^{\inte} \text{ and } e \in E^{\inte}}$.
\end{itemize}

We say that $\inte$ {\em models} an assertion $D \sqsubseteq E$ (resp. $R \sqsubseteq R'$),
and write $\inte \models D \sqsubseteq E$ (resp. $\inte \models R \sqsubseteq R'$),
whenever $D^{\inte} \subseteq E^{\inte}$ (resp. $R^{\inte} \subseteq R'^{\inte}$)
and a TBox
$\f{T}$, and write $\inte \models \f{T}$, whenever it is a model of all of its
assertions. We say that it {\em models} a membership assertion
$A(c)$ (resp. $R(c,c')$), and write $\inte \models A(a)$ (resp. $\inte \models R(c,c')$), 
whenever $c^{\inte} \in A^{\inte}$
(resp. $\tup{c^{\inte},c^{\inte}} \in R^{\inte}$) 
and an ABox $\f{A}$, and write $\inte \models \f{A}$,
when it {\em models} all of its membership assertions.
Finally, we say that it is a {\em model} of a KB $\tup{\f{T}, \f{A}}$,
and write $\inte \models \tup{\f{T}, \f{A}}$,
if it is a model of both $\f{T}$ and $\f{A}$.

The semantics \FOL formulas
is defined, we recall, in the usual terms
of satisfaction w.r.t. interpretations $\f{I}$.
Let $\phi$ be a \FOL formula and let
$\Var{\phi}$ denote the set of its variables.
An {\em assignment} 
for $\phi$ relative to $\inte$ is a
function $v \colon \Var{\phi} \to \domain$,
that can be recursively extended in the standard
way to complex formulas (see, e.g., \cite{CoriLascar}).
It is said to {\em satisfy} an atom $R(x_1,...,x_n)$ w.r.t. $\inte$
iff $\tup{v(x_1),...,v(x_n)} \in R^{\f{I}}$. This definition
is recursively extended
to complex formulas \cite{CoriLascar}. If $v$ satisfies $\phi$ w.r.t. $\inte$,
we write $\inte \models_v \phi$.
An interpretation $\inte$ is said to be a {\em model} of
$\phi$, written $\inte \models \phi$, 
if there exists an assignment $v$ s.t. $\inte \models_v \phi$.

A {\em union of conjunctive queries} (UCQ)
of arity $n$
is a (positive existential) \FOL formula of the form
$\phi := \psi_{1}(\vec{x},\vec{y}_1) \lor ... \lor 
\psi_{k}(\vec{x},\vec{y}_k)$ 
where 
$\vec{x}$ is a sequence of $n \geq 0$
{\em distinguished variables} and
the $\psi_i$s, for $i \in [1,k]$, are conjunctions of atoms.
A UCQ is said to be {\em boolean} if $\vec{x}$
is an empty sequence. The integer $\textit{size}(\phi)$
denotes the number of symbols of $\phi$.

Let $\tup{\f{T},\f{A}}$ be a KB and $\phi$ a UCQ of arity $n$. 
KB $\tup{\f{T},\f{A}}$ is said to {\em entail}
$\phi$, written 
$\tup{\f{T},\f{A}} \models \phi$,
iff for all interpretations $\inte$, $\inte \models \tup{\f{T},\f{A}}$
implies that $\inte \models \phi$.
The {\em certain answers} of a UCQ $\phi$ over KB 
$\tup{\f{T},\f{A}}$ are defined as the set
$\textit{cert}(q,\f{O},\f{D}) := \set{\vec{c} \in \e{Dom}^n \mid 
\f{T}, \f{A} \models \phi(\vec{c})}$, where
$\phi(\vec{c})$ denotes the instantiation of $\vec{x}$ in $\phi$
by a sequence of constants $\vec{c}$. The associated decision
problem is known as the KB {\em query answering} problem (\g{QA}) 
and is
defined as follows:
\begin{itemize}
\item given $\vec{c} \in \e{Dom}^n$, a UCQ $\phi$ of arity $n$ and a
KB $\tup{\f{T}, \f{A}}$,
\item does $\f{T}, \f{A} \models \phi(\vec{c})$?
\end{itemize}

When $\card{T}$ and $\textit{size}(\phi)$ are fixed we speak about the 
{\em data complexity} of \g{QA}, when only $\textit{size}(\phi)$ about its {\em KB complexity}, 
when $\card{T}$ and $\card{A}$ are fixed about its {\em query complexity}
and finally, when none is fixed, about its {\em combined complexity}.
It is known \cite{Calvanese2007C} that $\dlliterc$ is
in \LogSpace in data complexity, \PTime-complete in KB complexity and
\NP-complete in query complexity, but its combined complexity remains unknown.

%%%%%%%%%%%%%%%%%%%%%%%%%%%%%%%%%%%%%%%%%%%%%%%%%%%%%%%%%%%%%%%
\section{Combined Complexity of \g{QA}}\label{two}
%%%%%%%%%%%%%%%%%%%%%%%%%%%%%%%%%%%%%%%%%%%%%%%%%%%%%%%%%%%%%%%

A {\em perfect reformulation} is an algorithm
that takes as input a DL TBox $\f{T}$ and a UCQ $\phi$
and rewrites $\phi$ w.r.t. $\f{T}$ into a UCQ
$\phi_{\f{T}}$ s.t., for every DL ABox
$\f{A}$ and every $\vec{c} \in \e{Dom}$ it holds that:
$\f{T} , \f{A} \models \phi(\vec{c})$ iff $\f{I}(\f{A}) \models \phi_{\f{T}}(\vec{c})$,
where $\f{I}(\f{A})$ denotes the interpretation built out of $\f{A}$
(i.e., $\f{A}$ seen as a \FOL interpretation).

\begin{proposition}
{\bf (Calvanese et al. 2006)}
A perfect reformulation exists for $\dlliterc$.
\end{proposition}

\begin{theorem}
QA for $\dlliterc$ is \NP-complete in combined complexity.
\end{theorem}

\proof 
(\e{Membership}) Let $\tup{\f{T},\f{A}}$ be a
KB and let $\phi(\vec{c})$ be the grounding of a UCQ $\phi$.
First, consider:
$\f{T} , \f{A} \models \phi(\vec{c}).$
We know that
$\f{T}$ can be "compiled" into $\phi$ by a perfect reformulation, yielding a
UCQ
$\phi_{\f{T}}(\vec{c}) :=
\psi^{\f{T}}_1(\vec{c},\vec{y}_1) \lor ... \lor
\psi^{\f{T}}_k(\vec{c},\vec{y}_k)$. Guess, therefore, a disjunct
$\psi^{\f{T}}_i(\vec{c},\vec{y}_i)$, for some $i \in [1,k]$.
This can be done in time constant in $\card{\f{T}}$ and $\textit{size}(q)$.
Clearly, $\f{T} , \f{A} \models \phi(\vec{c})$ iff
$\f{I}(\f{A}) \models_v \psi^{\f{T}}_i(\vec{c},\vec{y_i})$, for
some assignment $v$. 
Guess now an assignment $v \colon \Var{\psi_i} \to 
\Delta^{\f{I}(\f{A})}$. 
This can be done in time constant in, ultimately, $\textit{size}(\phi)$.
Finally, check in
time polynomial on $\card{\f{A}}$ and $\textit{size}(\phi)$ whether
$\f{I}(\f{A})\models_{v} \psi_i(\vec{c},\vec{y}_i)$.

(\e{Hardness}) By reduction from the graph homomorphism problem, where, given
two graphs $G_1 = \tup{V_1, E_1}$ and $G_2 = \tup{V_2, E_2}$
we ask whether there exists an homomorphism $h$ from $G_1$ to $G_2$. 
A graph homomorphism, we recall, is a function $h \colon V_1 \to V_2$ s.t.
for all $\tup{u,v} \in V_1$, $\tup{h(u),h(v)} \in V_2$.
This problem is known to the \NP-complete.
We will consider $\dlliterc$ KBs with empty TBoxes.
Polynomially encode $G_1$ and $G_2$ as follows:
\begin{itemize}
\item for each $\langle u, v \rangle \in E_1$, add the fact $R(c_u,c_v)$
to the ABox $\f{A}_{G_1}$,
\item for each $\langle u', v' \rangle \in E_2$, add the ground atom $R(c_{u'},c_{v'})$
to the boolean UCQ $\phi_{G_2}$, which is the conjunction of such atoms.
\end{itemize}

We now claim that
there exists an homomorphism $h$ from graph $G_2$ to graph $G_1$
iff $\f{A}_{G_1} \models \phi_{G_2}$.

Since there is a perfect reformulation for $\dlliterc$, then
$\f{A}_{G_1} \models \phi_{G_2}$ iff $\f{I}(\f{A}_{G_1}) \models \phi_{G_2}$.
Now, clearly, $\f{I}(\f{A}_{G_1}) = G_1$.
Thus, the interpretation function $.^{\f{I}(\f{A}_{G_1})}$ can be seen as an homomorphism mapping
$\phi_{G_2}$ to $G_1$. Finally, given that $\phi_{G_2}$ encodes $G_2$, the claim
follows. \qed

%%%%%%%%%%%%%%%%%%%%%%%%%%%%%%%%%%%%%%%%%%%%%%%%%%%%%%%%%%%%%%%
\section{$\dlliterc$ Simulations}\label{three}
%%%%%%%%%%%%%%%%%%%%%%%%%%%%%%%%%%%%%%%%%%%%%%%%%%%%%%%%%%%%%%%

Given two interpretations $\f{I}$ and $\f{J}$, a 
{\em $\dlliterc$ left} $\f{B}_l$ or {\em right simulation} $\f{B}_r$ 
is a relation $\f{B}_l, \f{B}_r \subseteq \f{P}(\Delta^{\f{I}}) \times 
\Delta^{\f{J}}$ s.t., for every $X \subseteq \Delta^{\f{I}}$, every
$d' \in \Delta^{\f{J}}$\footnote{Observe that the clause for $D \sqcap D'$ follows implicitly from the first two.}:
\begin{itemize}
\item if $\tup{X,d'} \in \f{B}_l$ and $X \subseteq A^{\f{I}}$, then $d' \in \Delta^{\f{J}} \, (A)$.
\item if $\tup{X,d'} \in \f{B}_l$ and forall $d \in X$ there is some $e \in Y \subseteq \Delta^{\f{I}}$ 
s.t. $\tup{d,e} \in R^{\f{I}}$, then there exists an $e' \in \Delta^{\f{J}}$ s.t.
$\tup{d',e'} \in R^{\f{J}} \, (\exists R)$.
\item if $\tup{X,d'} \in \f{B}_r$ and $X \subseteq \neg B^{\f{I}}$, then $d' \not\in B^{\f{J}} \, (\neg B)$.
\item if $\tup{X,d'} \in \f{B}_r$ and forall $d \in X$ there exists no $e \in Y \subseteq \Delta^{\f{I}}$ s.t. 
$\tup{d,e} \in R^{\f{I}}$, then there is no $e' \in \Delta^{\f{J}}$ s.t.
$\tup{d',e'} \in R^{\f{J}} \, (\neg \exists R)$.
\item if $\tup{X,d'} \in \f{B}_r$ and forall $d \in X$ there exists an $e \in Y \subseteq \Delta^{\f{I}}$ s.t. 
$\tup{d,e} \in R^{\f{I}}$, then there is an $e' \in \Delta^{\f{J}}$ s.t.
$\tup{d',e'} \in R^{\f{J}}$ and $\tup{Y,e'} \in \f{B} \, (\exists R . C)$.
\end{itemize}

A {\em $\dlliterc$ simulation} $\f{B}$ is either a left, a
right or a combination of both simulations (i.e., their union). 
If a $\dlliterc$ simulation $\f{B}$ exists
among two interpretations ${\f{I}}$ and ${\f{J}}$ we say that
they are {\em DL-similar} and write ${\f{I}} \sim_{DL} {\f{J}}$.

We say that a \FOL formula $\phi$ is {\em closed under $\dlliterc$
simulations} iff for every two interpretations $\f{I}$ and $\f{J}$,
if $\f{I} \models \phi$ and $\f{I} \sim_{DL} \f{J}$, then $\f{J} \models \phi$.

We say that a \FOL formula $\phi$ {\em entails} 
a $\dlliterc$ concept $C$,
written $\phi \models C$,
iff for all $\inte$, $\inte \models \phi$ implies
that $C^{\inte} \neq \emptyset$, and conversely, that
$C$ {\em entails} $\phi$, written $C \models \phi$, whenever, for all
$\inte$, $C^{\inte} \neq \emptyset$ implies $\inte \models \phi$.
If both entailments hold, we say that they are {\em equivalent}.

\begin{lemma}
If A \FOL formula $\phi$ is closed
under \g{DL-Lite} simulations, then it is 
equivalent to a \g{DL-Lite}
right hand or left hand side concept.
\end{lemma}

\proof Let $\phi$ be a \g{FOL} formula closed under
$\dlliterc$ simulations. Let $\textit{Con}(\phi)$ denote the set of consequences in 
$\dlliterc$ of 
a \FOL formula $\phi$, i.e., $\textit{Con}(\phi) := \set{C \mid \phi \models C}$. 
By compactness for DLs \cite{DLHandbook} the set of concepts
$\textit{Con}(\phi)$ has a model iff every finite $\Sigma \subseteq \textit{Con}(\phi)$ 
has a model, whence
the concept $C_{\phi} := 
\bigsqcap \set{C \mid C \in \Sigma}$ should have a model too.
We claim that $\phi$ is equivalent to $C_{\phi}$. Clearly,
$\phi \models C_{\phi}$. We claim now that
\begin{equation}\label{eq:e}
C_{\phi} \models \phi.
\end{equation}
Assume that $C^{\inte}_{\phi} \neq \emptyset$, 
for an arbitrary intrepretation $\f{I}$.
Then, there exists a
$d \in \Delta^{\f{I}}$ s.t. $d \in C^{\inte}_{\phi}$.
Put now $\Gamma := \set{C \mid d \not\in 
C^{\f{I}}}$. Then, for every $C \in \Gamma, \phi \not\models C$. 
Hence for every $C \in \Gamma$
there exists an interpretation ${\f{I}}_C$ 
s.t. ${\f{I}}_C \models \phi$ and $C^{{\f{I}}_C}=\emptyset$.
The idea now is to build an interpretation $\f{J} := \tup{\Delta^{\f{J}},
.^{\f{J}}}$ from the
$\inte_C$s:
\begin{itemize}
\item $\Delta^{\f{J}} := \bigcup \set{\Delta^{\f{I}_C} \mid C \in \Gamma}$,
\item $.^{\f{J}}$ extends each $.^{\f{I}_C}$, for $C \in \Gamma$.
\end{itemize}
Define now a \g{DL-Lite} simulation $\f{B} \subseteq \f{P}(\Delta^{\f{J}}) \times \Delta^{\f{I}}$ by putting:
\begin{center}
\begin{tabular}{c}
$\tup{X,d'} \in \f{B}$ iff for every concept $C$, $X \subseteq C^{\f{J}}$ implies $d' \in C^{\f{I}}$.
\end{tabular}
\end{center}
We now claim that $\f{B}$ is a $\dlliterc$ simulation
between $\f{J}$ and $\f{I}$ and a fortiori that
$\f{J} \sim_{DL} \f{I}$.
We prove this by induction on $C$:
\begin{itemize}
\item Basis:
\begin{itemize}
\item The property trivially holds for basic concepts. 
\item $C := \neg A$. Let $X \subseteq \neg A^{\f{J}}$,
$\tup{X,d'} \in \f{B}$. By definition of $\f{B}$, $d' \in (\neg A)^{\f{I}}$, that is,
$d' \in \Delta^{\f{I}} - A^{\f{I}}$.
\item $C := \exists R$. Let $\tup{X,d'} \in \f{B}$
and $d \in X$ such that there is some $e \in Y \subseteq \Delta^{\f{J}}$
such that $\tup{d,e} \in R^{\f{J}}$. Now, $X \subseteq (\exists R)^{\f{J}}$, so
$d' \in (\exists R)^{\f{I}}$ and hence there is some $e' \in \Delta^{\f{I}}$
such that $\tup{d',e'} \in R^{\f{J}}$.
\item $C := \neg \exists R$. This is proven by combining the two
previous cases.
\end{itemize}
\item Inductive step:
\begin{itemize}
\item $C := \exists R . E$. Let $\tup{X,d'} \in \f{B}$ s.t.
exists $e \in Y \subseteq \Delta^{\f{J}}$ and
$\tup{d,e} \in R^{\f{J}}$. $X \subseteq (\exists R . E)^{\f{J}}$,
therefore, $d' \in (\exists R \colon D)^{\f{I}}$ by definition
and so there is an $e' \in \Delta^{\f{I}}$ such that $\tup{d',e'} \in R^{\f{I}}$
and $e' \in E^{\f{I}}$. Suppose that
$Y \subseteq E^{\f{J}}$. By induction hypothesis, $e' \in E^{\inte}$.
Thus, by definition of $\f{B}$,
$\tup{Y,e'} \in \f{B}$.
\item $C := D \sqcap D'$ (trivial).
\end{itemize}
\end{itemize}

Therefore, $\f{J} \sim_{DL} \f{I}$ 
and since by assumption $\phi$ is closed under
$\dlliterc$ simulations, $\f{I} \models \phi$.
This means that claim (\ref{eq:e}) holds. \qed

\begin{lemma}
If a \FOL formula $\phi$ is equivalent to a \g{DL-Lite}
right hand or left hand side concept, then it is closed
under $\dlliterc$ simulations.
\end{lemma}

\proof Let $\inte$ be s.t. $\inte \models \phi$. Let $\f{J}$ be an interpretation
DL-similar to $\inte$. Let
$X \subseteq \Delta^{\f{I}}, d \in X, d' \in \Delta^{\f{J}}, 
\f{B} \subseteq \f{P}(\Delta^{\f{I}}) \times \Delta^{\f{J}}$ and assume that
$\tup{X,d'} \in \f{B}$. 
We prove now, by induction on $C$, that $C^{\f{J}} \neq \emptyset$:
\begin{itemize}
\item Basis:
\begin{itemize}
\item $C := A$. Let $d \in A^{\f{I}}$.
Then, $X \subseteq A^{\f{I}}$, whence (by definition)
$d' \in A^{\f{J}}$.
\item $C := \neg A$ (analogous argument).
\item $C := \exists R$. Let
$d \in (\exists R)^{\f{I}}$. Then there exists
$e \in \Delta^{\f{I}}$ s.t. $\tup{d,e} \in R^{\f{I}}$, whence, by definition of
\g{DL-Lite} simulations $\f{B}$, 
there is an $e' \in \Delta^{\f{J}}$ s.t. $\tup{d',e'} \in R^{\f{J}}$, 
that is, s.t. $d' \in (\exists R)^{\f{J}}$.
\item $C := \neg \exists R$ (analogous argument).
\end{itemize}
\item Inductive step:
\begin{itemize}
\item $C := \exists R . E$. 
Suppose that $d \in (\exists R \colon E)^{\f{I}}$. Therefore there
is some $e \in \Delta^{\f{I}}$ s.t. $e \in E^{\f{I}}$ and $\tup{d,e'} \in R^{\f{I}}$.
By induction hypothesis this implies that $e \in E^{\f{J}}$, whence
$d \in (\exists R . E)^{\f{J}}$ as well.
\item $C := D \sqcap D'$. By induction hypothesis
the property holds for $D$ and $D'$. Now:
\begin{equation*}
\begin{array}{ccl}
d \in (D \sqcap D')^{\f{I}}   	& \text{ iff }     & d \in D^{\f{I}} \text{ and } d \in D'^{\f{I}}\\
		 	        & \text{ implies } & d' \in D^{\f{J}} \text{ and } d' \in D'^{\f{J}}\\
	                        & \text{ iff }     & d' \in (D \sqcap D')^{\f{J}}.
\end{array}
\end{equation*}
\end{itemize}
Therefore, since $\phi$ is equivalent to $C$,
$\f{J} \models \phi$, as desired. \qed
\end{itemize}

\begin{theorem}
A \FOL formula $\phi$ is equivalent to a $\dlliterc$
right hand or left hand side concept iff it is closed
under \g{DL-Lite} simulations.
\end{theorem}

\begin{example}
The \FOL formula $\phi := \forall y P(x,y) \to A(y)$ is not equivalent to any
$\dlliterc$ concept, because
it is not closed under $\dlliterc$ simulations.

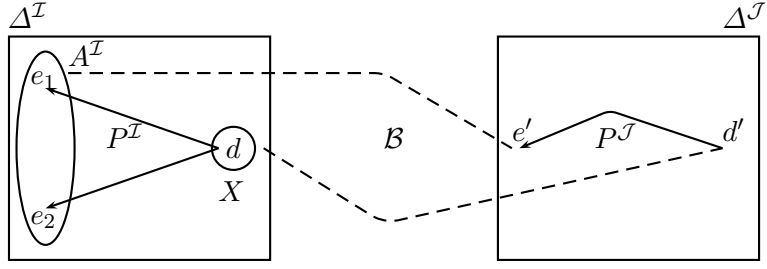
\begin{center}
\begin{pspicture}(10,4.5)
\psline{->}(2.8,2.5)(0.5,3.3)
\psline{->}(2.8,2.5)(0.5,1.7)
\psellipse(0.5,2.5)(0.4,1.3)
\put(0,4.1){$\Delta^{\f{I}}$}
\put(9.5,4.1){$\Delta^{\f{J}}$}
\put(5,2.5){$\f{B}$}
\put(1.3,2.5){$P^{\f{I}}$}
\put(7.8,2.5){$P^{\f{J}}$}
\put(0.8,3.6){$A^{\f{I}}$}
\cput[doubleline=false](3,2.5){$d$}
\put(2.8,1.8){$X$}
\put(0.3,3.35){$e_1$}
\put(0.3,1.5){$e_2$}
\put(6.7,2.6){$e'$}
\psline[linearc=.25]{<-}(6.8,2.5)(8,3)(9.5,2.5)
\put(9.5,2.6){$d'$}
\psframe(0,1)(3.5,4)
\psframe(6.5,1)(10,4)
\psline[linestyle=dashed,linearc=.5](0.8,3.5)(5,3.5)(6.7,2.5)
\psline[linestyle=dashed,linearc=.5](3.4,2.5)(5,1.5)(9.5,2.5)
\end{pspicture}
\end{center}
As the reader can see, $\f{B}$ is a $\dlliterc$ simulation
there \textit{(i)} $\tup{\set{d},d'} \in \f{B}$, \textit{(ii)} $\tup{\set{e_1,e_2},e'} \in \f{B}$
and \textit{(iii)} $\f{I} \sim_{DL} \f{J}$.
Now, clearly,
$\f{I} \models_{v[x:=d]} \forall y P(x,y) \to A(y)\}$, but
$\f{J} \not\models_{v'[x:=d']} \forall y P(x,y) \to A(y)$, since
$A^{\f{J}} = \emptyset$. \ex
\end{example}

%%%%%%%%%%%%%%%%%%%%%%%%%%%%%%%%%%%%%%%%%%%%%%%%%%%%%%%%%%%%%%%
\section{Some Negative Results}\label{four}
%%%%%%%%%%%%%%%%%%%%%%%%%%%%%%%%%%%%%%%%%%%%%%%%%%%%%%%%%%%%%%%

\begin{proposition}
Disjunction is not expressible in $\dlliterc$.
\end{proposition}

\proof $\dlliterc$ is contained in \g{HORN} 
(the set of \FOL horn clauses)\cite{Calvanese2007C,Calvanese2007E},
which cannot express disjunctions of the form
$\phi := A(c) \lor A'(c')$. 
Otherwise, let $\f{H} := \set{A(c)}$ 
and $\f{H'} := \set{A'(c')}$
be two Herbrand models of $\phi$. Clearly, $\f{H}$ and $\f{H'}$
are minimal (w.r.t. set inclusion) models of $\phi$ s.t. $\f{H} \neq \f{H'}$.
But this is impossible, since \g{HORN}
verifies the least (w.r.t. set inclusion) 
Herbrand model property \cite{CoriLascar}. \qed

\begin{theorem}
There is no relation $\sim$ over interpretations 
such that,
for every \FOL sentence $\phi$, $\phi$ is equivalent to a
$\dlliterc$ assertion iff it is closed under 
the relation $\sim$.
\end{theorem}

\proof Recall that a \FOL sentence is a 
\FOL formula with no free variables.
Suppose the contrary and consider the sentence
$A(c)$. Let $\f{I}$ and $\f{J}$ be two 
structures s.t. $\f{I} \sim \f{J}$ and suppose that
$\f{I} \models A(c)$. Then, obviously, $\f{J} \models A(c)$ too.
But then:
\begin{center}
\begin{tabular}{l}
$\f{I} \models A(c)$ implies $\f{I} \models A(c) \lor A'(c)$, and\\
$\f{J} \models A(c)$ implies $\f{J} \models A(c) \lor A'(c)$.
\end{tabular}
\end{center}
That is,  $A(c) \lor A'(c)$ is closed under $\sim$ and is a fortiori
equivalent to some
$\dlliterc$ assertion.
But this is impossible, because disjunction is not expressible
in $\dlliterc$. \qed

%%%%%%%%%%%%%%%%%%%%%%%%%%%%%%%%%%%%%%%%%%%%%%%%%%%%%%%%%%%%%%%
\section{Conclusions}\label{five}
%%%%%%%%%%%%%%%%%%%%%%%%%%%%%%%%%%%%%%%%%%%%%%%%%%%%%%%%%%%%%%%

In this paper we have shown four things:
\textit{(i)} Answering UCQs over $\dlliterc$ KBs is \NP-complete in
combined complexity.
\textit{(ii)} A simulation relation among interpretations, viz., a $\dlliterc$ simulation, 
can be used to characterize the expressive power of $\dlliterc$ as a
concept language.
\textit{(iii)} \FOL formulas that are closed under $\dlliterc$ simulations
are equivalent to a (left or right) $\dlliterc$ concept.
\textit{(iv)} This closure property holds only w.r.t. concepts, but not w.r.t.
assertions. 
Simulations, in particular, can be generalized, with minor adjustments, to
the whole \g{DL-Lite} family of DLs, although, since all of them are in
\g{HORN}, no such closure property exists for their assertions.

%%%%%%%%%%%%%%%%%%%%%%%%%%%%%%%%%%%%%%%%%%%%%%%%%%%%%%%%%%%%%%%
\bibliographystyle{plain}
\bibliography{/home/camilo/Documents/bibs/my_bib}
%%%%%%%%%%%%%%%%%%%%%%%%%%%%%%%%%%%%%%%%%%%%%%%%%%%%%%%%%%%%%%%

%%%%%%%%%%%%%%%%%%%%%%%%%%%%%%%%%%%%%%%%%%%%%%%%%%%%%%%%%%%%%%%
\end{document}